\documentclass[12pt,a4paper]{article}

\usepackage[utf8]{inputenc}
\usepackage[T1]{fontenc}
\usepackage{amsmath,amssymb,amsfonts}
\usepackage{graphicx}
\usepackage{textcomp}
\usepackage[margin=2.5cm]{geometry}
\usepackage{multirow}
\usepackage{xcolor}
\usepackage{cite}
 \usepackage{booktabs}
\usepackage{algorithm}
\usepackage{algorithmic}
\usepackage{listings}
\usepackage[utf8]{inputenc}
\usepackage[T1]{fontenc}
\usepackage{textcomp}
\usepackage{url} 

\usepackage{amsmath, amssymb, amsfonts}

\usepackage{multirow} 
\usepackage{array}    
\usepackage{booktabs} 

\usepackage{graphicx}

\usepackage{listings}
\usepackage{algorithm}
\usepackage{algorithmic}

\usepackage{cite}
\usepackage{xcolor}
\usepackage[hidelinks]{hyperref} 
\lstdefinelanguage{Solidity}{
    keywords={break, case, catch, continue, delete, do, else, enum, export, external, for, function, if, import, indexed, inheritance, interface, internal, library, mapping, modifier, new, payable, public, pragma, private, returns, return, selfdestruct, struct, throw, using, view, while, pure, constructor, event, emit},
    keywordstyle=\color{blue}\bfseries,
    ndkeywords={address, bool, byte, bytes, bytes1, bytes32, int, int256, uint, uint256, string, var},
    ndkeywordstyle=\color{teal}\bfseries,
    identifierstyle=\color{black},
    sensitive=true,
    comment=[l]{//},
    morecomment=[s]{/*}{*/},
    commentstyle=\color{gray}\ttfamily,
    stringstyle=\color{red}\ttfamily,
    morestring=[b]',
    morestring=[b]"
}

\title{SmartGraphical: A Human-in-the-Loop Framework for Detecting Smart Contract Logical Vulnerabilities via Pattern-Driven Static Analysis and Visual Abstraction}

\author{
  Ali Fattahdizaji*$^{1}$,
  Mohammad Pishdar$^{2}$,
  Zarina Shukur$^{3}$ \\
  \small $^{1}$Information Science and Technology Department \\
  \small Universiti Kebangsaan Malaysia, Kuala Lumpur, Malaysia \\
  \small Email: p127035@siswa.ukm.edu.my \\
  \small $^{2}$CoinFa Blockchain Research Lab, Tehran, Iran \\
  \small Email: m.pishdar@coinfa.top \\
  \small $^{3}$Information Science and Technology Department \\
  \small Universiti Kebangsaan Malaysia, Kuala Lumpur, Malaysia \\
  \small Email: zarinashukur@ukm.edu.my \\
  \small Corresponding author: Ali Fattahdizaji (e-mail: p127035@siswa.ukm.edu.my)
}

\date{}

\begin{document}

\maketitle

\begin{abstract}
Smart contracts are fundamental components of blockchain ecosystems; however, their security remains a critical concern due to inherent vulnerabilities. While existing detection methodologies are predominantly syntax-oriented, targeting reentrancy and arithmetic errors, they often overlook logical flaws arising from defective business logic. This paper introduces SmartGraphical, a novel security framework specifically engineered to identify logical attack surfaces. By synthesizing automated static analysis with an interactive graphical representation of contract architectures, SmartGraphical facilitates a comprehensive inspection of a contract's functional control flow. To mitigate the context-dependent nature of logical bugs, the tool adopts a human-in-the-loop approach, empowering developers to interpret heuristic warnings within a visualized structural context. The efficacy of SmartGraphical was validated through a rigorous empirical evaluation involving a large dataset of real-world contracts and a large-scale user study with 100 developers of varying expertise. Furthermore, the framework's performance was demonstrated through case studies on high-profile exploits, such as the SYFI rebase failure and farming protocol flash swap attacks, proving that SmartGraphical identifies intricate vulnerabilities that elude state-of-the-art automated detectors. Our findings indicate that this hybrid methodology significantly enhances the interpretability and detection rate of non-trivial logical security threats in smart contracts.

\end{abstract}


\section{Introduction}
\label{sec:introduction}

The growing use case of blockchain technology is mainly fueled by its inherent properties: decentralized consensus, immutability through cryptography, and persistent availability. In this setup, smart contracts—self-executing, programmable, and autonomous agents—are used to enable complex decentralized applications (dApps) and digital asset management. However, the "code is law" model suggests that any security gap or design weakness in the deployed contract can result in catastrophic and irreversible financial damage. Thus, the need to ensure the strong security of smart contract execution has become a pressing research need \cite{wen2023security,zou2019smart,alqahtani2023survey}.

Smart contracts are prone to a wide range of attacks, from syntax-level issues such as reentrancy and integer overflow attacks to design-level inconsistencies. Although state-of-the-art automated code analysis tools such as Oyente, Mythril, and Securify have shown strong effectiveness in detecting syntax-level vulnerabilities, they are largely ineffective in detecting logical attacks. Unlike syntax-level errors, which are deterministic and amenable to pattern matching, logical vulnerabilities are inherently context-dependent and arise from weaknesses in the underlying contract logic. The recent spate of high-profile attacks on contract logic suggests a pressing research gap: the absence of specialized tools that can bridge the gap between automated static analysis and the logical nuances of contract logic \cite{zou2019smart,wang2018overview}

To overcome this challenge, we propose SmartGraphical, a new hybrid approach specifically designed to fill the gap between automated detection and human reasoning. Unlike traditional detectors, SmartGraphical enables a multi-faceted analysis through the following features:

\begin{itemize}
    \item Logical Relationship Mapping: Uncovering and displaying the complex functional relationships and dependencies between the various contract elements.
    \item Heuristic-based Alerting: Automated alerting of high-quality warnings that point to possible logical discrepancies.
    \item Human-in-the-Loop Workflow: A semi-automated verification step where developers analyze automated warnings in the context of the visualized structural representation of the contract.

\end{itemize}
To assess the efficacy of our approach, we designed an extensive empirical assessment. First, we assessed the detection accuracy of the tool on 100 real-world smart contracts. Second, we conducted a multi-dimensional usability assessment with 100 developers with varying levels of expertise. Unlike previous assessments, which solely depended on automated metrics, our assessment approach illustrates that the combination of human reasoning and graphical visualization can effectively decrease the "false-positive" rate and improve the interpretability of complex logical errors.

The main contributions of this work are:

\begin{itemize}
    
    \item The design of SmartGraphical, an open-source platform that combines static analysis with graphical inspection.
    \item Heuristic algorithms tailored for the automated identification of logical security inconsistencies.
    \item A two-fold assessment using real-world smart contract data and a large-scale user study (n=100) to assess the practical usefulness of the tool in real-world auditing settings.
\end{itemize}

The rest of this research is structured as follows: Section 2 gives definitions of logical security threats and background. Section 3 describes the methodology of this research. Section 4 introduces the related work about smart contract security. Section 5 describes the architectural design and visual semantics of the SmartGraphical framework. The modules and heuristic warning algorithms are described in Sections 6 to 9. Finally, Section 10 introduces the experimental results, developer feedback, and case studies of real-world logical attacks.

\section{BACKGROUND}
In this section, we will look at the bare minimum of information required to comprehend the research.

\subsection{Blockchain as a State Transition Machine }
Instead of being a simple distributed database, a blockchain is formally specified as a replicated deterministic state machine. It preserves the integrity of the network via a series of cryptographically chained blocks, where each block holds a series of atomic transactions that move the network from one legal state to another \cite{zou2019smart,wang2018overview}. The heart of this system is the State Transition Function (STF), which guarantees that all nodes, no matter how geographically distributed, agree on a single global state via a distributed consensus protocol.

Regarding smart contract security, the consensus algorithm (such as Proof of Work or Proof of Stake) is the finality layer, which guarantees that once a state transition (such as a contract execution) is verified and written to the ledger, it becomes fixed and immutable. In the realm of logical vulnerability analysis, this immutability is a double-edged sword: while it guarantees the integrity of execution, it also makes any exploited logic irreversible. Thus, the blockchain system requires a "correct-by-construction" paradigm, where the functional logic must be formally verified before execution, since the underlying execution environment (such as the Ethereum Virtual Machine) will faithfully execute malicious business logic as accurately as it would secure code.\cite{singh2020blockchain,mourtzis2023blockchain}.

\subsection{Smart Contract Architecture and Execution Paradigm}

Smart contracts are deterministic, self-executing programs running on decentralized ledgers, most commonly known as the Ethereum Virtual Machine (EVM). Contrary to being mere legal contracts, they are State Transition Systems, in which the "code is law" paradigm regulates the relationship between users and digital assets [2, 4]. The functional soundness of these contracts relies on the harmonious interplay of five essential architectural elements \cite{dannen2017solidity,buterin2014next,documentation2021solidity,solidity2024}:

\begin{itemize}
    \item Persistent State Variables: These are the contract’s own memory, symbolizing the unalterable data repository that embodies the system’s state during the course of transactional executions.
    \item Executable Functions: These encode the contract’s business logic and enable user interactions, allowing for asset transfer and state variable manipulation.
    \item Conditional Modifiers: These are modular gatekeepers that enforce security preconditions (such as access control) and extend functional behavior.
    \item Event-Driven Logging: Notification systems that trigger signals during state transitions, offering a critical audit trail for external observation.
    \item External Libraries: Code repositories that promote modularity and extend the contract’s computational capabilities
\end{itemize}

The structural paradigm of such components is generally coded in Solidity, a high-level, contract-centric language that draws on C++ and Python, specifically tailored for the EVM and compatible blockchains like Binance Smart Chain \cite{dannen2017solidity,documentation2021solidity}. In order to better understand the complex interdependencies at play in this paradigm, the following formal definition of the anatomy of a contract can be used: \\

\begin{lstlisting}[language=Solidity, caption={Structural Anatomy of a Smart Contract}, label={lst:example1}]

// Simplified paradigm of state-logic interaction
contract SmartParadigm {
    address public owner;
    uint256 public totalSupply;
    
    event Transfer(address indexed from, uint256 value);
    
    modifier onlyOwner() {
        require(msg.sender == owner, "Unauthorized");
        _;
    }
    
    function updateSupply(uint256 newValue) public onlyOwner {
        totalSupply = newValue;
        emit Transfer(msg.sender, newValue);
    }
}
\end{lstlisting} 

As shown in Listing 1, the internal dynamics are described by a set of interlinked dependencies where executable procedures interact with persistent data within the boundaries of access control protocols. To better understand the described relationships within a practical example, we analyze a Decentralized Auction Mechanism (see Listing 2). This example illustrates the interaction between state variables (highestBid), functions (placeBid), and events (NewBid). \\

In this design, the execution flows are determined by the state of the contract. The complexity in this design is where the logical vulnerabilities exist. A bug in the business logic (such as not refunding the previous bidder in Listing 2) could result in catastrophic financial damage, even if the code is syntactically correct \cite{dannen2017introducing,fattahdizaji2024investigating}.

 \begin{lstlisting}[language=Solidity, caption={Algorithmic Logic of an Auction Contract}, label={lst:example1}]
contract SimpleAuction {
    address public highestBidder;
    uint256 public highestBid;
    event NewBid(address bidder, uint256 amount);

    function placeBid() public payable {
        require(msg.value > highestBid, "Bid too low");
        highestBidder = msg.sender;
        highestBid = msg.value;
        emit NewBid(msg.sender, msg.value);
    }
}
\end{lstlisting} 
\subsection{Logical Attack with example}

Applications that contain logical flaws are a particular kind of issue that can occasionally turn into security risks. These dangers are not specific code flaws; rather, they are contingent on the nature of the business and the tool's architecture. Put another way, the application functions well and generates its output, but it may experience issues and exhibit unusual behavior in certain unique and unanticipated conditions that the application developer has not anticipated \cite{defi_platform}. Cybercriminals may detect these kinds of issues and attempt to subject the program to unexpected circumstances in order to obtain the intended result. In certain situations, this can be extremely risky and seriously harm the application's reputation, assets, or user data. Take a look at the following sample for a clearer understanding. The value of the total supply in a real smart contract is calculated as follows. This calculation is not problematic for the compiler and even works correctly in many cases in the real environment, and the end-user has no problems \cite{defi_platform}.
\\ \\
 $$totalSupply = initSupply.mul(yamsScalingFactor);$$
\\ \\
This is the correct form of this calculation, which should be as follows because it should be limited to the maximum possible value.
\\ \\
$$totalSupply = initSupply.mul(yamsScalingFactor).div(BASE);$$
\
\\ \\
The problem with logical vulnerabilities is that they are inherently dangerous because they are able to evade traditional detection methods; that is, since they do not cause runtime errors or compilation errors, they are often latent in traditional quality assurance processes. From the perspective of the user, the contract will seem to be fully functional. However, because the source code of smart contracts is typically public and immutable, it serves as a transparent roadmap for attackers to follow in order to exploit the subtle differences. A cyber attacker can use these vulnerabilities in order to manipulate key state variables, such as the total supply value, in order to disrupt the economic balance of the contract or cause a complete denial of service.
\section{Research Approach and Experimental Design}
In order to evaluate the SmartGraphical framework and enhance its heuristic detection capabilities, this research work adopts a multi-step approach that progresses from the theoretical foundation to empirical validation. This approach to research methodology begins with a literature review to identify the current knowledge gap in the existing tools, and then relies on the expert analysis of real-world exploit reports to build a high-fidelity dataset of logical vulnerabilities. The collected data and knowledge were then leveraged to build a dedicated pattern detection tool for smart contract source code. By combining this automated detection capability with a 'Human-in-the-Loop' (HITL) validation process, we can ensure that the framework is grounded in real-world attack patterns and has a high degree of accuracy in detecting complex business logic flaws that are normally out of the reach of automated scanners.

\subsection{Preliminary Systematic Review}

In our earlier published work [12], we undertook a taxonomy and longitudinal study of all blockchain attack vectors based on a carefully curated dataset of real-world breach reports. One of the main aims of this work was to close the existing gap between vulnerabilities reported in the literature and those that are actually seen in the wild. By carefully examining the data for the frequency and severity of these real-world breaches, we found that there was a profound mismatch: whereas traditional syntax- and property-based approaches such as symbolic execution and fuzzing are highly effective at mitigating standard vulnerabilities, they are, in fact, fundamentally unsuitable for modeling the complex logical dependencies that underlie the most common types of attacks.

The results obtained in [12] indicated that logic-related anomalies not only occur frequently but are also the most frequent source of serious financial damage, while they constitute a crucial ‘blind spot’ in state-of-the-art defensive tools. The above-mentioned empirical results directly influenced the creation of SmartGraphical. Based on the taxonomies and vulnerability patterns developed in our previous research work [12], this study begins with a thorough assessment of the current state-of-the-art solutions to identify their particular weaknesses in identifying logical anomalies. Our assessment shows that the current automated solutions are not context-aware enough to intercept complex business logic attacks. Thus, instead of just pointing out the weaknesses, we proceed to design SmartGraphical, graphical, pattern-based static analysis framework particularly designed to fill the identified gaps. The tool will be designed to display and identify the complex context-aware attack surfaces that have been shown to be the most persistent threats in the Ethereum network.

\subsection{Dataset Acquisition and Empirical Sampling}

The research started with the systematic gathering of historical smart contract security incidents. Rather than using random sampling, we carefully curated a dataset by examining actual attack reports from reputable blockchain security sources \cite{fattahdizaji2024investigating,defi_platform,defillama_hacks}. Two security professionals manually reviewed these reports to select and filter vulnerabilities that were specifically linked to complex business logic errors in Solidity smart contracts.

After the expert analysis, we obtained the source code for each exploit found directly from the Ethereum network. This step led to the creation of a high-quality benchmark dataset with 100 vulnerable smart contracts, each of which had been found to have actual logical errors. This dataset formed the basis of the ground truth from which vulnerability patterns were extracted and the SmartGraphical framework detection algorithms were tuned in the next phase.

\subsubsection{Research Questions (RQs)}
Motivated by the need to fill the gap between automated static analysis and human insight, the following research questions are answered:

\begin{itemize}
    \item RQ1 (Characterization): What are the distinguishing structural and behavioral features of logical vulnerabilities in deployed smart contracts?
    \item RQ2 (Visualization Efficacy): To what extent does graphical abstraction alleviate the cognitive burden of auditors in tracing functional dependencies?
    \item RQ3 (Heuristic Precision): How well do the proposed automated warning systems alert developers to subtle logical attack surfaces?
    \item RQ4 (Usability): How well does the framework support different levels of developer expertise with respect to accuracy and usability?
\end{itemize}

\section{RELATED WORKS}
Existing work on smart contract security is mainly centered on the automated identification of syntax-level vulnerabilities. The identification of logical flaws is still an uncharted area. We classify the existing work into three paradigms:

\subsection{Formal Verification and Model Checking}

The early approaches relied on symbolic execution and model checking to validate contract properties. Researchers in \cite{shishkin2019debugging} and \cite{osterland2020model} proposed approaches to translate Solidity code into formal models (PROMELA/SPIN) to validate the execution logic. Although these approaches are formal, they are prone to state-space explosion and are more expert-intensive to specify correctness properties, which makes them less suitable for spotting intricate logical errors in different business contexts \cite{mavridou2019verisolid}, \cite{nelaturu2022correct}.

\subsection{Static and Dynamic Analysis}
The standard set of tools such as Oyente, Mythril, and Securify use static analysis to identify predefined patterns of vulnerabilities such as reentrancy and overflow attacks \cite{di2019survey,wohrer2018smart}. While these tools are efficient, they are "context-blind" and lack the ability to understand the intended business logic of a smart contract, thereby missing vulnerabilities even if the code is syntactically correct but logically flawed. Runtime analysis tools such as ContractLarva \cite{azzopardi2018monitoring} have tried to address this issue by monitoring execution, but these tools are exhibit limited efficacy in pre-deployment logic validation in analyzing logical exploits. It is also important to note that SmartGraphical targets functional logic inconsistencies and not external fraud schemes such as Rug Pulls, which may involve social engineering or control mechanisms as opposed to code-level inconsistencies. While deep learning-based methods \cite{bresil2025deep} improve pattern detection, they do not offer sufficient interpretability for logic verification, which SmartGraphical helps address through visual abstraction.

\subsection{Graphical Representation and Machine Learning}
To improve code understanding, some works \cite{harer2019comparison,pierro2021smart} analyzed the graphical extraction of smart contract data. Nevertheless, these graphical representations were mainly created for documentation purposes and not for security analysis. Likewise, machine learning-based methods \cite{gao2024sguard+,ray2020security,chen2023smart} are based on known vulnerability reports and do not have the required data to detect new logical risks. \\chen2023smart

As shown in Table 1, SmartGraphical is the only solution that can fill these gaps. Unlike other solutions, our solution combines the following:

\begin{itemize}
    \item Logical Attack Vector Detection: Designed specifically for non-syntactic vulnerabilities.
    \item Visual Dependency Modeling: Translating difficult code into a graphical format that can be managed.
    \item Human-in-the-Loop Integration: Allowing developers to verify automated warnings through a transparent verification process.
\end{itemize}

Our solution fills the "context-blindness" gap of traditional solutions, providing a more comprehensive security analysis, which will be explained in the following sections.

\begin{table}[ht]
\centering
\caption{Comparison of existing tools with respect to logical attacks and human-in-the-loop support}
\label{tab:comparison}
\begin{tabular}{|l|c|c|}
\hline
\textbf{Tool / Approach} & \textbf{Logical Attacks} & \textbf{Human-in-the-loop} \\ \hline
Oyente \cite{di2019survey,oyente}         & $\times$ & $\times$ \\ \hline
Mythril \cite{mythril}        & $\times$ & $\times$ \\ \hline
Securify \cite{tsankov2018securify}       & $\times$ & $\times$ \\ \hline
VeriSolid  \cite{mavridou2019verisolid}   & $\times$ & $\times$ \\ \hline
ContractLarva \cite{azzopardi2018monitoring}   & Partial  & $\times$ \\ \hline
\textbf{SmartGraphical (This work)} & \checkmark & \checkmark \\ \hline
\end{tabular}
\end{table}

\section{The SmartGraphical Framework: Architectural Design and Visual Semantics}
On the basis of empirical evidence provided by real-world exploits (Section 4), it is concluded that logical errors can arise from obscured functional dependencies. To overcome this, SmartGraphical abstracts the source code into a multi-dimensional graphical model, which helps the developer to detect irregular state transitions that are commonly missed by automated scanners.

\subsection{Smart Contract Component Identification}
The framework starts by analyzing the Solidity source code to extract the fundamental architectural building blocks. The taxonomy, based on standard specifications \cite{dannen2017solidity}, includes the following:

\begin{itemize}
    \item Structural Units: Extraction of Contracts, Libraries, and Interfaces.
    \item Execution Logic: Functions (Loops, Events, Call Conditions), Constructors, and System Calls (DelegateCall, Send, Transfer).
    \item State and Data: Data types, State Variables, and Event Emitters.
\end{itemize}

\subsection{Visual Dependency Modeling (Formalizing Connections)}
The essence of SmartGraphical innovation is in its capability to translate code-level interactions into a directed graph, with nodes denoting components and arrows symbolizing functional dependencies between them. The meaning of these relationships is as follows:

\begin{itemize}
    \item Data Access Flow (Function $\to$ Variable): An arrow from a function node to a state variable represents a read/write action. This is essential for state integrity auditing.
    \item Input Dependency (Variable $\to$ Function): This indicates data structures necessary for the execution of a function, which can be exploited by malicious inputs.
    \item Initialization Logic (Constructor $\to$ Function/Variable): This represents the initialization step of the contract. Unauthenticated re-initialization or incorrect constructor logic is a common source of logical vulnerabilities.
    \item Systemic Interactions: The contract’s boundary is explicitly represented by special arrows, which symbolize low-level calls and interactions with other contracts.
    \item Control Flow Granularity: The internal logic of for/while loops and conditional statements is represented in geometric nodes to represent complexity and logical bottlenecks.
\end{itemize}

\subsection{Operational Workflow: From Code to Heuristic Warnings}
The working of SmartGraphical is a systematic process as shown in Figure 10:

\begin{itemize}
    \item Static Analysis Phase: The tool takes in the Solidity code, identifying the structural elements and their relationships.
    \item Graphical Synthesis: A graphical representation (for example, the SimpleAuction contract \cite{pishdar_simpleauction_2025} in Figure 2, and Visual Graph Components in Figure 1) is created to give a high-level view of the functional flow of the contract.
    \item Heuristic Alerting Phase: the extracted metadata for signatures of known logical attack surfaces. These signatures are derived from the high-frequency vulnerability patterns identified through our empirical analysis of real-world exploits. The system matches the contract’s functional dependencies against these predefined patterns to detect anomalies such as unauthorized state transitions, inconsistent access control, and incorrect business workflows. To aid in this process, the developer is offered a Command-Line Interface (CLI) (as depicted in Figure 3), which acts as the control center for running a particular audit task. Once a possible threat is identified, the system does not stop at giving a warning notification; instead, it activates the visual modeling engine to graph the dubious logic, enabling the auditor to correlate the automated results with a visual representation for final 'Human-in-the-Loop' validation.
    \item Human-in-the-Loop (HITL) Verification: The developer, by analyzing the graphical representation (for example, multiple unauthorized functions accessing the same data structure), can logically deduce "illogical" associations that would otherwise be identified as syntactically correct by an automated system.
\end{itemize}

\begin{figure}
    \centering
    \includegraphics[width=5cm,height=8cm,keepaspectratio]{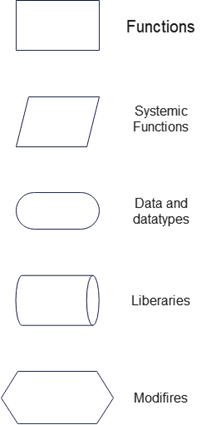}
    \caption{Elements that are utilized in visual depencency modeling}
    \label{fig:enter-label}
\end{figure}

\begin{figure*}
    \centering
    \includegraphics[width=1\linewidth]{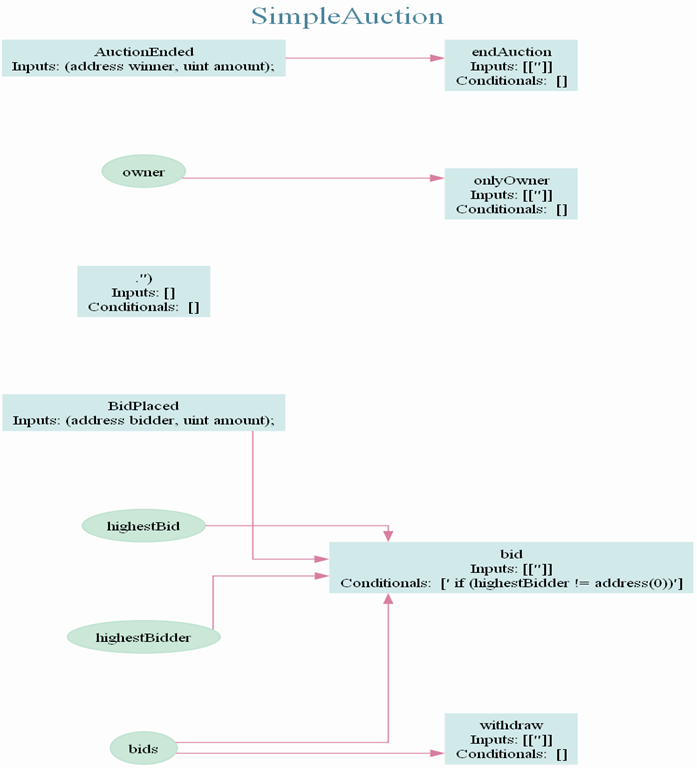}
    \caption{The Smart Graphical tool's output for SimpleAuction Smart Contract}
    \label{fig:enter-label}
\end{figure*}

\begin{figure*}
    \centering
    \includegraphics[width=1\linewidth]{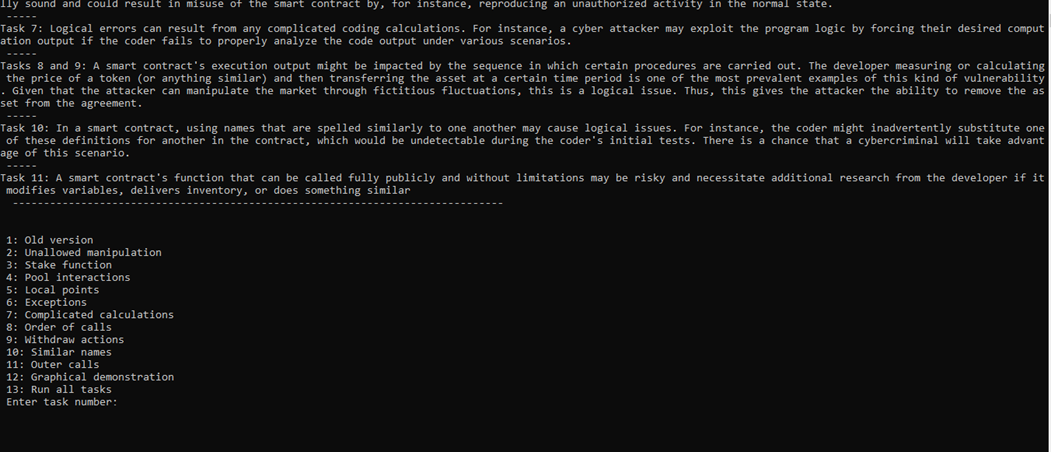}
    \caption{Command-Line Output of Logical Security Issues}
    \label{fig:enter-label}
\end{figure*}

This graphical abstraction serves as a diagnostic tool for identifying external influences on critical variables, such as supply values or dynamic fees. By visualizing the interaction between external inputs and internal state changes, SmartGraphical alerts the developer to variables that must not be treated as static, thereby preventing catastrophic logical failures. The final implementation of the framework is available as an open-source project
\cite{pishdar_smartgraphical_2025}.

\section{Module-Based Architecture for Feature Extraction}
The initial phase of SmartGraphical entails a systematic decomposition of the source code of the Solidity contract. As shown in Figure 4, the framework employs a modular parsing engine that is capable of extracting the critical metadata of the contract via pattern-based static analysis. Each module is tasked with a particular semantic level of the smart contract:

\begin{itemize}
    \item Contract and Interface Extraction: These modules extract individual smart contracts and their respective APIs (Interfaces) in a single .sol file. This ensures that in multi-contract files, each entity is analyzed in a logical, isolated context.
    \item Constructor and Modifier Parsing: Unlike regular functions, these modules parse initialization code (Constructors) and behavioral constraints (Modifiers). Modifier extraction is essential for access control logic analysis, which is a common source of logical bugs.
    \item Data Structure Analysis (Variables and Structs): These modules break down state variables (numeric, string, etc.) and complex data structures (Structs). By doing so, the tool can later analyze how data flows through the system.
    \item Functional Logic Engine (Functions, Loops, and Conditionals): This is the central module that identifies the function bodies, including internal control flow statements such as loops (for/while) and conditional statements (if/else). These statements are the basic units of the graphical model.
    \item Dependency Mapping (Variable-Function Mapping): This is a unique module that defines the relationships between the state variables and the functions that interact with or change them. This is crucial for visualizing the Data-Flow Graph.
    \item External Integration (Imports and Events): These functions are responsible for the identification of external dependencies and event emitters. Events are important logs that are used for off-chain validation and are modeled to represent the contract’s interaction with the blockchain ledger.
\end{itemize}

\begin{figure*}
    \centering
    \includegraphics[width=1\linewidth]{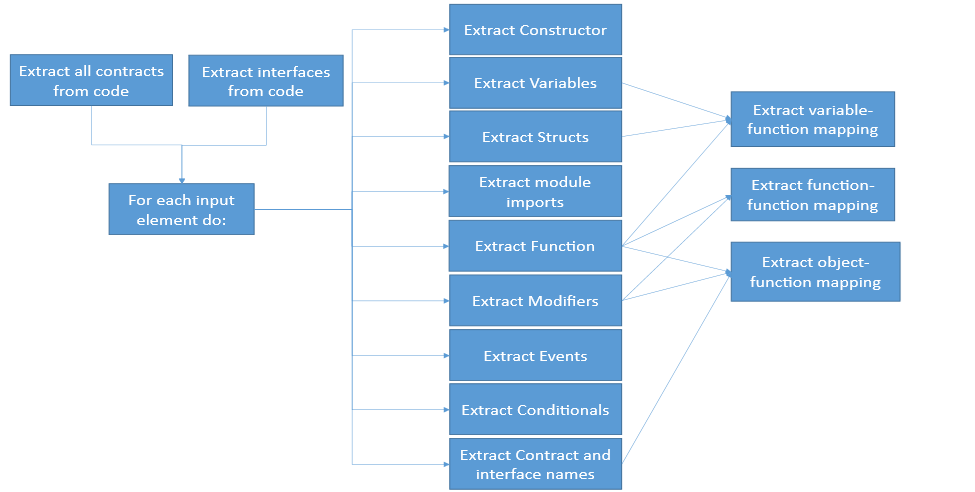}
    \caption{Various modules are employed, along with the connections among them}
    \label{fig:enter-label}
\end{figure*}

\section{Heuristic Warning Generation and Logical Indicators}
Although graphical modeling is a great improvement in terms of code understandability, the identification of subtle logical errors needs another level of automated support. SmartGraphical incorporates a set of heuristic audit modules to detect "red flags" or "indicators of logical inconsistency." These are not conclusive vulnerabilities but rather Heuristic Alerts that require "Human-in-the-Loop" verification (See Figure 5).

Based on the analysis of 104 verified exploits from our previous work [12], we formalized eleven main logical risk categories into the detection engine of the tool:

\subsection{External Dependency and State Integrity}
\begin{itemize}
    \item External Oracle/Contract Dependency: The heuristic triggers when critical parameters (such as totalSupply, maxFee, or price feeds) are calculated from external contracts without adequate checks and balances for range validation and variation.
    \item Supply Manipulation Hooks: Code that has arbitrary burns or mints is checked to ensure that these actions are rationally consistent with the pool calculations. 
    \item Collateral and Debt Logic (DeFi): In the DeFi space, the heuristic checks the interaction between stake, unstake, and collateral release to prevent unauthorized borrowing or collateral-less withdrawals.
\end{itemize}

\begin{figure*}
    \centering
    \includegraphics[width=1\linewidth]{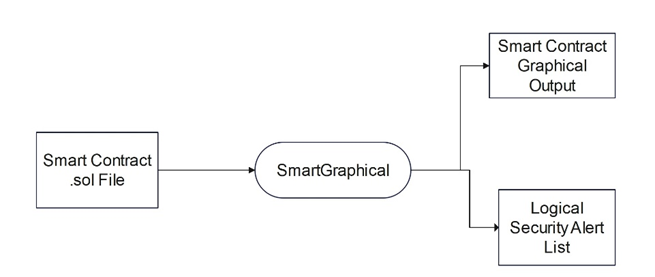}
    \caption{SmartGraphical Input/output Diagram}
    \label{fig:enter-label}
\end{figure*}

\subsection{Transactional and Economic Logic}
\begin{itemize}
    \item Balance-Logic Discrepancies: When a contract enables the deposit of assets (tokens, points, or staking), the tool points out discrepancies in the withdrawal logic to prevent users from withdrawing more than their entitled balance.
    \item Incentive Alignment and Point Systems: When a contract has internal reward points, the tool checks the "Earn-to-Spend" consistency to prevent points from being spent without being earned.
    \item Price Manipulation and Execution Ordering: Heuristics identify discrepancies in which asset transfers are based on highly volatile or manipulable time-windows to prevent attacks such as "sandwich attacks" or price manipulation.
\end{itemize}

\subsection{Computational and Operational Flaws}
\begin{itemize}
    \item Complex Calculation Risks: Any complex mathematical calculation involving multiple state transitions is highlighted for manual audit to avoid the possibility of tampering with the results by malicious actors.
    \item Non-Deterministic State Recovery: Issues related to non-deterministic system function calls, such as low-level calls, are detected, with special emphasis on the logical validity of state reversions (reverts) or the possibility of ending up in an inconsistent system state.
    \item Unrestricted Public Access: Public functions that alter critical state variables or inventory without access restriction modifiers (such as onlyOwner) are marked for high-priority alerts.
\end{itemize}

\subsection{Semantic and Maintenance Errors}
\begin{itemize}
    \item Legacy Signature Mismatches: For contracts that are a rewrite or an update, the tool points out function signatures that could potentially and unwittingly inherit outdated logic or parameters from the previous versions.
    \item Naming Ambiguity (Homoglyph/Shadowing): The tool points out variables or functions that have a high degree of similarity in naming, such as “owner” and “owners,” which could potentially result in unwitting substitution and logical errors.
\end{itemize}

\section{Formalization of Logic Detection Algorithms}
In order to systematically search for the logical indicators described in Section 7, we have developed a set of specialized algorithms for detection. It is important to note that the algorithms developed are of the Heuristic Classifier type; they have a high recall rate in order to detect subtle anomalies, which are then shown to the developer for manual validation through the graphical interface. The logical risk categories, as described in our taxonomy, have been assigned a particular algorithmic implementation in the SmartGraphical framework. This section describes the formal logic for four detection scenarios.

We have created certain algorithms to analyze the code and produce warnings when a logical error is likely to occur in order to assist the developer in identifying the issues brought up in the preceding section. It goes without saying that the creation of warnings does not necessarily indicate a serious issue; instead, the programmer must investigate the situation further by looking at the code and using tools like graphical output. Every logical mistake definition from the previous part has been translated into at least one algorithm and put into practice. The Smart Graphical code with the appropriate labels displays the implementations. This section presents pseudocode for the algorithm used in four of the definitions in the previous section. 
	
\subsection{Heuristic for Economic Inconsistency}
As shown in Algorithm 1, the algorithm aims at irregularities in asset logic (stake, deposit, or points). The detection engine traces the execution in a multi-step process:

\begin{enumerate}
    \item Keyword Semantic Mapping: The engine identifies functions and state variables related to asset management (stake, deposit, balance).
    \item State Mutation Tracing: The engine traces the execution order of operations on the variables.
    \item Logical Flow Validation: The algorithm looks for a "Check-Effect-Interaction" or "Balance Consistency" pattern. If the execution order of state mutations versus asset transfers does not look normal or lacks boundary values, the system marks the function as "Suspicious Status," which triggers a visual inspection by the developer.
\end{enumerate}

\begin{algorithm}
\caption{Detection of Asymmetric Logical Manipulations in Staking Workflows}
\label{alg:staking_detection}
\begin{algorithmic}[1]
\STATE \textbf{Initialize:} $Stake\_Possible\_Names \leftarrow$ ["stake", "deposit", "lock", "provide"]
\STATE \textbf{Initialize:} $Unstake\_Possible\_Names \leftarrow$ ["unstake", "withdraw", "unlock", "release"]
\STATE \textbf{Extract} all functions $F$ from the contract
\STATE \textbf{Identify} $F_{stake} \subset F$ where function name $\in Stake\_Possible\_Names$
\STATE \textbf{Identify} $F_{unstake} \subset F$ where function name $\in Unstake\_Possible\_Names$

\FORALL{function $f \in F_{stake}$}
    \STATE \textbf{Extract} all state variables $V_{stake}$ modified within $f$
    \IF{$V_{stake}$ is not empty}
        \FORALL{function $g \in F_{unstake}$}
            \STATE \textbf{Extract} all state variables $V_{unstake}$ modified within $g$
            
            \STATE \COMMENT{Check for Asymmetry: Stake to Unstake}
            \IF{$\exists v \in V_{stake} \mid v \notin V_{unstake}$}
                \STATE \textbf{Flag:} "Inconsistent State Update: Missing Unstake Logic for $v$"
            \ENDIF
            
            \STATE \COMMENT{Check for Asymmetry: Unstake to Stake}
            \IF{$\exists u \in V_{unstake} \mid u \notin V_{stake}$}
                \STATE \textbf{Flag:} "Inconsistent State Update: Missing Stake Logic for $u$"
            \ENDIF
        \ENDFOR
    \ENDIF
\ENDFOR
\STATE \textbf{Return} All identified anomalies and alerts
\end{algorithmic}
\end{algorithm}

\subsection{Heuristic for Staking Logic and Sequential Execution}
As shown in Algorithm 2, the framework uses a sequential flow analysis to audit staking and unstaking functions. The algorithm is intended to guarantee that state transitions are safeguarded by proper preconditions. The steps include:

\begin{itemize}
    \item Identifier Extraction: The engine searches the contract for semantic identifiers associated with staking functionality (e.g., stake, unstake, claim).
    \item Execution Trace: It traces all low-level system calls and state modifications corresponding to the identifiers.
    \item Linear Flow Analysis: Based on the execution line numbers, the analyzer traces the sequence of instructions.
    \item Precondition Validation: The algorithm focuses on the presence of conditional validation (e.g., require or assert statements) before the functions. In case a critical state transition happens without a validated precondition, a Logic-Sequence Warning is raised.
\end{itemize}

\begin{algorithm}
\caption{Validation Check Analysis for Financial Exit Functions}
\label{alg:validation_check}
\begin{algorithmic}[1]
\STATE \textbf{Initialize:} $Exit\_Keywords \leftarrow$ ["withdraw", "unstake", "claim", "redeem", "transfer"]
\STATE \textbf{Extract} all functions $F$ and state variables from the contract
\STATE \textbf{Identify} $F_{exit} \subset F$ where function name contains any keyword $\in Exit\_Keywords$

\FORALL{function $f \in F_{exit}$}
    \STATE \textbf{Analyze} control flow and internal calls of $f$
    \STATE \textbf{Extract} all nested/systematic function calls $f_{sys}$ invoked within $f$
    
    \STATE \COMMENT{Analyze Validation Mechanisms}
    \STATE $Found\_Check \leftarrow \text{False}$
    
    \IF{$f$ contains any \texttt{require(...)} statement}
        \STATE $Found\_Check \leftarrow \text{True}$
    \ELSIF{$f$ contains any \texttt{if(...)} conditional check}
        \STATE $Found\_Check \leftarrow \text{True}$
    \ELSIF{$f$ contains any \texttt{try-catch} or \texttt{except} block}
        \STATE $Found\_Check \leftarrow \text{True}$
    \ENDIF
    
    \STATE \COMMENT{Alert Generation for Missing Logic}
    \IF{$Found\_Check$ is \text{False}}
        \STATE \textbf{Generate Alert:} "High Risk: Missing Validation Logic in function $f$"
        \STATE \textbf{Include} $f_{sys}$ in the audit report for context analysis
    \ENDIF
\ENDFOR
\STATE \textbf{Return} Summary of all identified unprotected exit paths
\end{algorithmic}
\end{algorithm}

\begin{algorithm}
\caption{Detection of Unprotected External Entry Points and Parameter Manipulation}
\label{alg:unprotected_functions}
\begin{algorithmic}[1]
\STATE \textbf{Extract} all function definitions $F$ and inheritance graph $G$ from the contract
\STATE \textbf{Identify} all internal call-sites $C$ within the contract body

\FORALL{function $f \in F$}
    \STATE \COMMENT{Step 1: Identify independent/external entry points}
    \IF{$f$ is not invoked by any internal function $\in C$}
        \STATE \textbf{Check} inheritance: Is $f$ called by parent or child contracts in $G$?
        
        \IF{$f$ is not called by related contracts \AND $f.visibility == \text{External}$}
            
            \STATE \COMMENT{Step 2: Check for missing Access Control}
            \IF{$f$ does not possess any modifier containing 'only' (e.g., \textit{onlyOwner}, \textit{onlyRole})}
                
                \STATE \COMMENT{Step 3: Data Flow Analysis of Input Parameters}
                \STATE \textbf{Extract} input parameters $P_f$ of function $f$
                
                \FORALL{parameter $p \in P_f$}
                    \IF{$p$ is used to set or manipulate state variables or contract logic}
                        \STATE \textbf{Generate Alert:} "Potential Vulnerability: Unprotected external function $f$ manipulates state via param $p$"
                    \ENDIF
                \ENDFOR
                
            \ENDIF
        \ENDIF
    \ENDIF
\ENDFOR
\STATE \textbf{Return} List of high-risk entry points and flagged parameters
\end{algorithmic}
\end{algorithm}

\subsection{Heuristic for Unrestricted Public Access}
Algorithm 3 aims at pinpointing high-risk entry points that are not protected by sufficient access control.

\begin{itemize}
    \item Visibility Audit: Pinpointing functions with public or external visibility that are not called internally from other contract components.
    \item Impact Analysis: The algorithm analyzes whether these functions interact with critical contract variables or perform complex arithmetic operations on external inputs.
    \item Warning Generation: When a function has high-impact functionality (such as state modification) and unrestricted public visibility without proper modifiers, it is highlighted as a "Suspicious Entry Point" for urgent manual analysis.
\end{itemize}

\subsection{Heuristic for Price Manipulation and Execution Ordering}
As formalized in Algorithm 4, SmartGraphical tackles the important problem of temporal dependencies in asset pricing. This heuristic is tailored to identify potential weaknesses in which the state can be manipulated between a price query and an asset transfer. The detection logic follows a three-step protocol:

\begin{algorithm}
\caption{Detection of Execution Order Anomalies in Price-Dependent Transfers}
\label{alg:price_transfer_order}
\begin{algorithmic}[1]
\STATE \textbf{Initialize:} $Price\_Keywords \leftarrow$ ["rebase", "fetchPrice", "updatePrice", "getRate"]
\STATE \textbf{Initialize:} $Transfer\_Keywords \leftarrow$ ["transfer", "transferFrom", "safeTransfer", "send"]
\STATE \textbf{Threshold:} $D_{max} \leftarrow$ Maximum allowable code distance (e.g., number of operations)

\FORALL{function $f$ in the contract}
    \STATE \textbf{Identify} all call sites $C_{price} \subset f$ matching $Price\_Keywords$
    \STATE \textbf{Identify} all call sites $C_{transfer} \subset f$ matching $Transfer\_Keywords$
    
    \IF{$C_{price} \neq \emptyset$ \AND $C_{transfer} \neq \emptyset$}
        \FORALL{pair $(cp, ct)$ where $cp \in C_{price}$ and $ct \in C_{transfer}$}
            \STATE \textbf{Calculate} $Distance \leftarrow$ lines of code or operations between $cp$ and $ct$
            
            \STATE \COMMENT{Analyze Execution Sequence and Gap}
            \IF{$Distance > D_{max}$ \OR $ct$ precedes $cp$}
                \STATE \textbf{Generate Alert:} "Potential Price-Lag Vulnerability: Excessive logic gap between price update and transfer in $f$"
                \STATE \textbf{Flag} intermediate operations for "Flash Loan" or "Price Manipulation" risk
            \ENDIF
        \ENDFOR
    \ENDIF
\ENDFOR
\STATE \textbf{Return} Set of flagged functions with inconsistent call sequences
\end{algorithmic}
\end{algorithm}

\begin{itemize}
    \item Functional Extraction: The engine looks for all calls related to price oracles and asset valuation, as well as functions related to asset transfers or state-dependent trades.
    \item Sequential Inter-dependency Mapping: By analyzing the line numbers of execution and the operational flow of the contract, the tool determines the relative order of these calls.
    \item Vulnerability Trigger: If the algorithm finds that the price query and the subsequent transfer are decoupled, in the sense that an attacker might have the ability to manipulate the price or state variables between these two points, it produces a High-Priority Logical Warning. This is intended to prompt the developer to add atomicity or further slippage checks
\end{itemize}

\subsection{Syntactic vs. Logical Detection: A Comparative Paradigm}
Contrary to conventional security solutions (e.g., Mythril, Oyente) that mainly focus on syntax-level issues like reentrancy and arithmetic overflow, the alerts produced by SmartGraphical are based on high-level logical patterns that have been identified through empirical analysis of exploit data.

Each alert is more than a generic error message and represents a heuristic attack on specific design flaws, including:

\begin{itemize}
    \item Economic Inconsistencies: Mismanagement of token distribution and collateral unlocking.
    \item Access Control Anomalies: Public calls without restrictions that alter the critical state.
    \item Operational Sequencing: Business logic flaws that enable temporal attacks.
\end{itemize}

Through the combination of these new heuristics with a human-in-the-loop graphical interface, SmartGraphical fills the existing gap between automated symbolic execution and static analysis tools and instead targets the business logic flaws that are most common in contemporary smart contract attacks.

\section{EVALUATION}
In order to assess the efficacy and usability of the SmartGraphical framework, we conducted a comprehensive user study with a total of 100 professional developers. For this purpose, participants were recruited from a wide variety of global and regional freelance ecosystems, including but not limited to Upwork.com, Parscoders.com, Karlancer.com, etc. This ensured that a wide spectrum of technical backgrounds and expertise was represented among the participants. Detailed demographic information about the participants' proficiency in general programming, blockchain security, smart contract development, logical exploit analysis, etc., is presented in Figure 6. In order to accommodate the diverse skill sets of the participants, the participants were
grouped according to their level of familiarity with the domain (Table 2). All the experimental datasets and contracts are available at \cite{Pishdar:2025github}.

The evaluation process was conducted through a series of two distinct phases. Firstly, in order to create a performance baseline, we asked the developers to audit four different smart contracts that contained undisclosed logical flaws without the use of our tool. Figure 8 presents a graphical representation of their baseline performance in detecting logical flaws. Secondly, we evaluated the efficacy of the SmartGraphical tool by asking the developers to audit two complex smart contracts that were derived from actual hacked protocols. These contracts contained "historically exploited logical flaws" as opposed to "injected logical flaws."

As shown in Figure 7, the overall pool of participants initially had a relatively low level of foundational knowledge in terms of smart contract logic security, but the addition of SmartGraphical greatly improved the diagnostic capabilities of the overall pool.

A comparative analysis of the overall audit results, as shown in Figure 8, revealed that without the aid of the SmartGraphical tool, only an infinitesimal percentage of developers, even those with high levels of expertise in related technical fields, had the ability to recognize the logical security vulnerabilities that had been integrated into the code. However, the overall addition of the warning system through the use of the SmartGraphical tool created a paradigm shift in the overall performance, as shown in Figure 9. The overall recognition of logical security vulnerabilities improved dramatically, but the overall trend was particularly pronounced in terms of the overall performance of highly skilled developers, as shown in Figure 9.

The diagnostic reliability of SmartGraphical was quantitatively measured based on a dataset of 100 smart contracts. In order to test the precision and sensitivity of the tool in a balanced environment, the dataset was comprised of both verified vulnerable and non-vulnerable smart contracts. The performance metrics, including the rates of correct security warnings and intentional false alarms, are presented in Table 3. As indicated in the table, although the tool has a high rate of recall to ensure that no logical nuances are overlooked, the hybrid 'human-in-the-loop' architecture of the tool enables the efficient filtering of alarms.

The data provided on the rate of invalid detection by developers is a vital piece of information on a crucial part of the SmartGraphical ecosystem. Although the automated engine is highly sensitive to ensure maximum coverage, as provided by the tool, the process of reviewing it manually by developers (as depicted in Figure 10) is the final step for validating the code. The rate of invalid detection is a reflection of the 'safety first' heuristic approach provided by the tool to ensure maximum scrutiny by the human auditor. The results have shown that while the system is capable of identifying a wide range of suspicious patterns, human intuition is vital to differentiate between complex logical vulnerabilities and code structures to refine the precision of the tool.

\begin{table}
\centering
\caption{Categorizing developers by level in each field
}
\begin{tabular}{| p{2cm} | p{10cm} |}
\hline
The level & Level of developer in every field of familiarity with programming, coding security, smart contracts, and logical attacks

  \\
\hline
Zero Knowledge & Not passing specialized courses and relevant work experience \\
\hline
Low & Passing specialized courses without relevant work experience or real projects \\
\hline
Middle & Having less than 3 years of work experience \\
\hline
High & Having more than 3 years of work experience and passing relevant specialized courses \\
\hline

\end{tabular}

\end{table}

\begin{table}[h]
\centering
\caption{Quantitative Performance Evaluation of SmartGraphical Diagnostic Engine}
\begin{tabular}{|l|l|c|c|c|}
\hline
\textbf{Category} & \textbf{Metric} & \textbf{Value 1} & \textbf{Value 2} & \textbf{Result} \\ \hline
\multirow{3}{*}{Dataset Composition} & Smart Contracts Analyzed & 500 & 300 & 800 \\ \cline{2-5} 
                                     & Verified Vulnerable & 45 & 20 & 65 \\ \cline{2-5} 
                                     & Non-vulnerable & 455 & 280 & 735 \\ \hline
\multirow{2}{*}{Detection Performance} & Sensitivity (Recall) & 92\% & 89\% & 91\% \\ \cline{2-5} 
                                       & Specificity & 95\% & 94\% & 94.5\% \\ \hline
\end{tabular}
\end{table}

\begin{figure*}
    \centering
    \includegraphics[width=12cm,height=4cm]{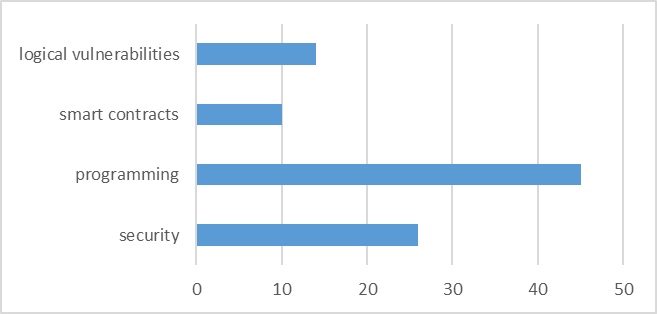}
    \caption{people with upper middle knowledge in different fields}
    \label{fig:enter-label}
\end{figure*}

\begin{figure*}
    \centering
    \includegraphics[width=13cm,height=6cm]{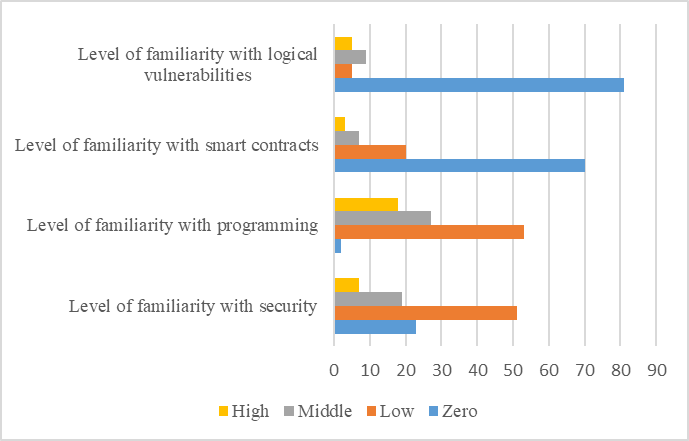}
    \caption{The degree to which developers are knowledgeable about various topics}
    \label{fig:enter-label}
\end{figure*}

\begin{figure*}
    \centering
    \includegraphics[width=15cm,height=6cm]{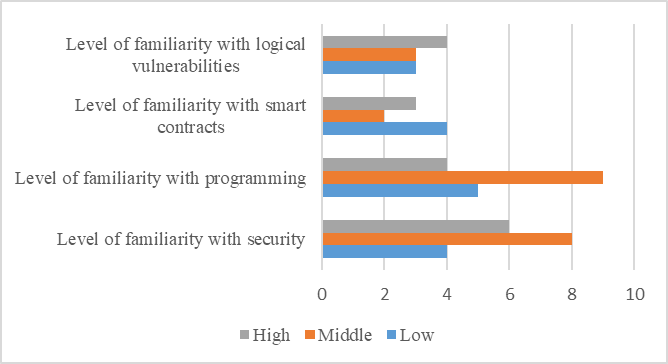}
    \caption{The detection rate of logical security flaws without using Smart Graphical warnings}
    \label{fig:enter-label}
\end{figure*}

\begin{figure*}
    \centering
    \includegraphics[width=13cm,height=6cm]{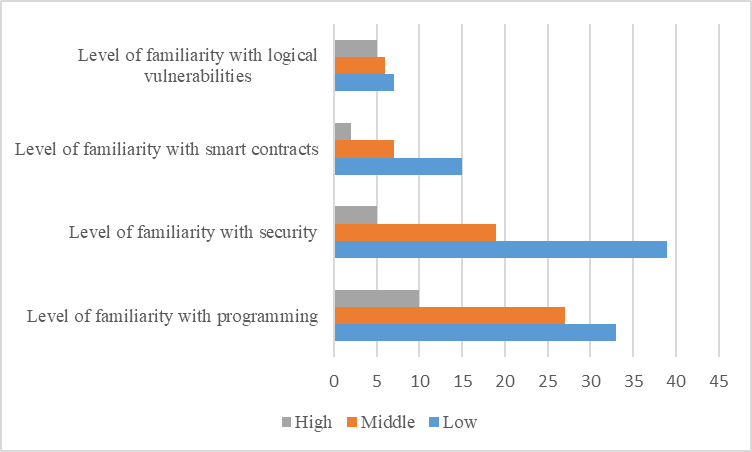}
    \caption{The detection rate of logical security flaws using Smart Graphical warnings}
    \label{fig:enter-label}
\end{figure*}

\begin{figure*}
    \centering
    \includegraphics[width=13cm,height=6cm]{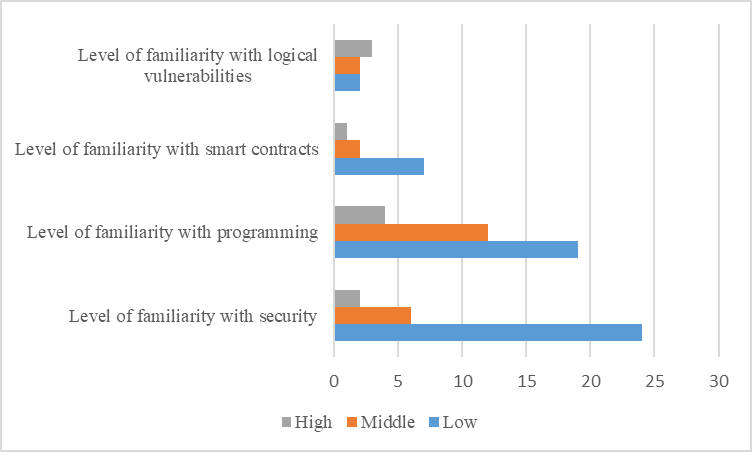}
    \caption{the rate of developers' invalid detection using Smart Graphical}
    \label{fig:enter-label}
\end{figure*}

In order to further affirm the effectiveness of the SmartGraphical approach, we performed an additional comparative study with two of the best-performing Large Language Models (LLMs) in the market, namely, ChatGPT-4o and Gemini 1.5 Pro. These models were presented with the source code of the same set of verified vulnerable contracts, as used in the initial dataset, i.e., the ones categorized by the 'Logical' type of vulnerabilities in our reverse-engineered dataset. The objective was to have the models perform the identification of business logic vulnerabilities without any assistance or abstraction. As shown in Table 3, the LLMs were found to have performed at an average baseline in the identification of security patterns, but they performed poorly in the identification of logical vulnerabilities, as compared to the performance of the SmartGraphical approach, where the average detection rate was found to be 85.7\%, while the LLMs performed at an average of 48\%.

\begin{table}[htbp]
\centering
\caption{Comparative Analysis: Detection Rate of Real-World Logical Vulnerabilities}
\label{tab:llm_comparison}
\begin{tabular}{|l|c|c|c|}
\hline
\textbf{Vulnerability Category} & \textbf{ChatGPT-4o} & \textbf{Gemini 1.5 Pro} & \textbf{SmartGraphical} \\ \hline
Business Logic Errors & 43\% & 45\% & \textbf{85\%} \\ \hline
Price Oracle Manipulation & 37\% & 32\% & \textbf{78\%} \\ \hline
Reward Calculation Errors & 51\% & 55\% & \textbf{88\%} \\ \hline
\textbf{Average Detection Rate} & \textbf{43.6\%} & \textbf{44\%} & \textbf{85.7\%} \\ \hline
\end{tabular}
\end{table}

\section{Case Studies on Real-world Logical Attacks}
To further confirm that SmartGraphical is good at finding logical vulnerabilities, we tested it on two real-world smart contracts that had already been hacked. These contracts are examples of high-impact situations where logical design flaws, not coding mistakes, caused huge financial losses. The following case studies show how SmartGraphical was able to find suspicious logical patterns and send out warnings that other tools couldn't.

\subsection{Example 1 – SYFI Rebase Attack}
The rebase mechanism wasn't set up right, which caused a big logical problem with the SYFI protocol (Table 3). The protocol allowed people to start a Uniswap sell transaction right after they changed the balance of their wallet, but before any changes in the price of tokens were made. The wrong conversion rate between YFI and SYFI also caused the wrong changes to the balance. These problems let an attacker steal almost all of the pool's money.
Type of vulnerability: not correctly handling rebase and not keeping state consistent.
SmartGraphical found: A warning about doing things in the wrong order and getting balance updates that don't make sense.
They can't find this because these tools (Oyente, Mythril, Securify) only look for arithmetic or reentrancy errors. \\

\begin{table}[htbp]
\centering
\caption{Analysis of the SYFI Protocol Logical Exploit}
\label{tab:syfi_analysis}
\resizebox{\textwidth}{!}{ 
\begin{tabular}{|l|p{6cm}|p{6cm}|} 
\hline
\textbf{Category and Detail} & \textbf{Description} & \textbf{Significance} \\ \hline
Protocol Name & SYFI (Soft Yearn Finance) & DeFi Rebasing Token \\ \hline
Attack Vector & Positive Rebasing Manipulation & Logical Logic Drift \\ \hline
Initial Position & 0.5 ETH $\rightarrow$ 2 SYFI & Seed Asset Acquisition \\ \hline
Exploit Transaction & 0x853...c4e68e & Balance Amplification \\ \hline
Liquidation Transaction & 0xb15...ae33 & Immediate Arbitrage \\ \hline
Net Profit & 747 ETH & High-Impact Exploit \\ \hline
\end{tabular}
}
\end{table}

\subsection{Example 2 – FORM Farming Flash Swap Exploit}
The USD value of LP and FORM tokens and rewards was figured out using pool balances in one of the farming contracts (Table 4). Because this contract used an unsecured price discovery mechanism, it could be changed by a Flash Swap. The attacker changed the price of the LP position by making it look higher so they could get more rewards. This vulnerability was an unsafe reward calculation logic caused by an unsafe pool balance dependency. SmartGraphical identified this as a warning that there is an unsafe external dependency utilized in price assessment and reward calculation, yet existing tools could not detect it since they were unable to model such a scenario. \\

\begin{table}[htbp]
\centering
\caption{Technical Specification of the Cross-Chain Logical Exploit}
\label{tab:multichain_exploit}
\begin{tabular}{@{}ll@{}}
\toprule
\textbf{Entity} and \textbf{Identifier / Address} \\ \midrule
Main Attack Transaction and \texttt{0x5c9a...b6ab} (Ethereum) \\
Adversarial Contract & \texttt{0xb5ae...711e} \\
Affected Contract (BSC) & \texttt{0xe2ee...0012} \\
Affected Contract (ETH) & \texttt{0x6293...334b} \\
\bottomrule
\end{tabular}
\end{table}

These case studies demonstrate that SmartGraphical can identify logical errors arising from design and business logic, rather than from syntactical errors in the code. In both cases, the tool
gave warnings about logical structures that looked suspicious. This helped developers
figure out what was causing the problems. Oyente, Mythril, and Securify are examples
of automated detectors that are already available. They couldn’t find these problems
because they mostly look for errors like reentrancy, arithmetic errors, or access control problems.
This evaluation confirms the uniqueness and effectiveness of our approach in addressing
actual logical attacks.
\section{Limitations and Threats to Validity}
In this section, we recognize some limitations that may affect the generalizability and performance of the SmartGraphical framework:

\begin{itemize}
    \item Internal Validity (Heuristic Accuracy): As the SmartGraphical framework uses a set of heuristic patterns defined through Regex, its accuracy may be limited by the quality and extent of these heuristics. For instance, complex contract structures may not be detected. To address this, we made the SmartGraphical framework extensible, allowing users to add new heuristics as new attack vectors emerge.
    \item External Validity (User Expertise): As a Human-in-the-Loop system, the accuracy of the SmartGraphical framework may depend on the user's level of expertise. A user who is not familiar with the system may misinterpret the abstracted information presented in the graphical interface. To address this, we conducted a user study with 100 participants to ensure that the system's graphical user interface was user-friendly for users with varying skill levels. 
    \item Construct Validity (Scope of Detection): The SmartGraphical framework targets functional logic vulnerabilities. It does not detect off-chain attacks such as Rug Pulls or private key thefts, which do not involve the execution logic of a smart contract.
\end{itemize}

\section{CONCLUSION and Future Work}
This study proposed SmartGraphical, a hybrid security solution aimed at filling the crucial gap in logical vulnerability detection in smart contracts. By integrating automated static analysis with graphical modeling, the solution offers a multi-layered defense mechanism that goes beyond the conventional syntax-based security solution.

Our empirical study, based on the analysis of 139 real-world exploits, confirmed that logical errors are the leading cause of high-impact breaches in the Ethereum network. The performance analysis of SmartGraphical on a benchmark set of 100 vulnerable smart contracts showed its high effectiveness, with a 89\% true positive detection rate for logical errors. Although the tool produced heuristic alerts in 82\% of the cases, the findings clearly highlight the need for our Human-in-the-Loop (HITL) solution. The addition of visual dependency mapping greatly improved the performance of the developers, increasing the detection rate of intricate design errors from 12\% to 58.3\%.
The results verify the importance of the synergy between automated heuristic notifications and graphical abstraction in the context of context-dependent logical bug detection. Unlike conventional scanners, SmartGraphical enables auditors to understand functional interdependencies, which remain invisible to fully automated systems.

Future work will concentrate on the following major tracks:

\begin{enumerate}
    \item Algorithmic Optimization: Incorporating machine learning algorithms to improve the accuracy of heuristic detection and minimize the false positive rate.
    \item Scalability: Extending the framework to other leading blockchain platforms, aside from Ethereum, such as Binance Smart Chain and Polygon.
    \item Comprehensive Taxonomy: Incorporating additional logical attack patterns, including governance attacks and cross-chain bridge attacks.
\end{enumerate}

SmartGraphical marks a critical milestone in the direction of a more resilient and developer-friendly security model in the ever-changing decentralized application ecosystem.

\bibliographystyle{ieeetr} 
\bibliography{References} 

\section*{Statements and Declarations}
\subsection*{Funding}
This research did not receive any specific grant from funding agencies in the public, commercial, or nonprofit sectors. The authors and Universiti Kebangsaan Malaysia, Kuala Lumpur, Malaysia, supported this work.

\subsection*{Competing Interests
}
The authors declare that they have no financial or non-financial competing interests to disclose. During the preparation of this work, the authors used Grammarly to improve the text. After using this tool, the authors reviewed and edited the content as needed and take full responsibility for the content of the published article. The authors would also like to thank the assistance of AI-based grammar and style checkers, which were applied sparingly to improve the readability of the manuscript.

\subsection*{Author Contributions
}Ali Fattahdizaji contributed to the study conception, design, and analysis of the results. Mohammad Pishdar was responsible for the initial research, data collection, and writing of the first draft of the manuscript. Zarina Shukur provided supervision and critical review. All authors have reviewed and approved the final manuscript.

\end{document}